\documentclass{elsart}
\usepackage{natbib}
\usepackage{graphicx}
\begin{document}
\runauthor{Meynet et al.}
\begin{frontmatter}
\title{$^{26}$Al yields from rotating Wolf--Rayet star models}
\author[La]{Christel Vuissoz}
\author[Ge]{Georges Meynet}
\author[To]{Jurgen Kn\"odlseder}
\author[Sp]{Miguel Cervi\~no}
\author[Ge]{Daniel Schaerer}
\author[Br]{Ana Palacios}
\author[Is]{Nami Mowlavi}

\address[La]{Institut d'astronomie de l'Universit\'e de Lausanne, 1290 Chavannes-des-Bois, Switzerland}
\address[Ge]{Geneva Observatory, CH--1290 Sauverny, Switzerland}
\address[To]{Centre d'Etude Spatiale des Rayonnements, CNRS/UPS, BP 4346, 31028 Toulouse Cedex 4, France}
\address[Sp]{Instituto de Astrof\'{\i}sica de Andaluc{\'{\i}}a (CSIC), Camino bajo de Hu\'etor 24, Apdo. 3004, 18080 Granada, Spain}
\address[Br]{Institut d'Astronomie et d'Astrophysique, Universit\'e Libre de Bruxelles, CP 226, 1050 Brussels, Belgium}
\address[Is]{INTEGRAL Science Data Center, Chemin d' Ecogia 16, CH--1290 Versoix, Switzerland}

\begin{abstract}
We present new $^{26}$Al stellar yields from rotating Wolf--Rayet stellar models
which, at solar metallicity, well reproduce the observed properties of the Wolf-Rayet
populations.
These new yields are enhanced with respect to non--rotating models,
even with respect to non--rotating models computed with enhanced mass loss rates.
We briefly discuss some implications of the use of these new yields for estimating
the global contribution of Wolf-Rayet
stars to the quantity of $^{26}$Al now present in the Milky Way.
\end{abstract}
\end{frontmatter}

\section{Introduction}

Many papers, in the past, have explored the possibility 
that Wolf-Rayet stars,  through their winds, might enrich the
interstellar medium in $^{26}$Al. 
The idea was first suggested by Dearborn and Blake \cite{ref2} and
was further explored by many authors
(see the review by Prantzos and Diehl \cite{ref3} and references therein). 
These computations showed
that indeed Wolf-Rayet stars appear to be significant $^{26}$Al sources.
It is the hope that in the future, $\gamma$--ray astronomy, by measuring the 1.8 MeV luminosity of
single objects, will be able to add new constraints on the stellar models and thus
clearly identify the source of this element in the Galaxy. An upper limit for the 1.8 MeV luminosity
exists only in the case
of  $\gamma^2$Vel, the nearest known WR star (Oberlack et al. \cite{ref4}). All the other constraints
concern either the Galaxy as a whole or directions in the Galaxy
with $gamma$--ray line  luminosities
arising from the cumulative effects of different populations whose relative contributions
are still difficult to assess. There is however a few very young associations or groups
of associations which are too young for having been much enriched
by the supernovae. In those cases, a significant part of the 1.8 MeV luminosity likely originates from
the decay of $^{26}$Al ejected by the winds of Wolf--Rayet (WR) stars. 
These regions are thus particularly interesting as they offer a unique probe
of one single type of sources (Kn\"odlseder et al. \cite{ref5}; Pl\"uschke et al. \cite{ref6}). 

A striking conclusion of the work
by Kn\"odlseder et al. \cite{ref5} who studied the Cygnus associations, 
is that the theoretical $^{26}$Al yields seem to be underestimated
by a factor of two. 
At first sight, this conclusion is surprising because
the WR yields are based on models computed with enhanced, 
(nowadays considered to be overestimated) mass loss rates (Meynet et al. \cite{ref7}).
One would have thus expected that the too 
strong mass loss suffered by these models
would give too high $^{26}$Al yields !
On the other hand, despite the uncertainties
pertaining to the mass loss rates, there was good hope that the yields of these
models would not be too far from reality in the sense that the enhanced
mass loss rate models were able to account for many observable properties
of WR stars as for instance  the number ratio of WR to O type stars in zones of constant star formation rate 
(Maeder \& Meynet \cite{ref8}). 

Recently, a new grid of WR stellar models at solar metallicity has been computed accounting
for the effects of rotation and including new prescriptions for the mass loss rates.
The new revised mass loss rates include the effects due to clumping in WR stellar winds
and are about a factor two to three below those
used in the enhanced mass loss rate models. 
Interestingly, 
despite using much lower mass loss rates,
the rotating models can explain the observed number ratio
of WR to O-type stars at solar metallicity. They can even
reproduce the observed fraction 
of WR stars presenting at their surface
both H and He-burning products, an observational fact which
non--rotating models could not explain (with normal
or enhanced mass loss rates). Finally they also
well match the number ratio of WC to WN stars.
Thus it appears that the necessity to enhance the mass loss rate in the old models
was probably due to the neglect of rotation. 

The aim of this paper is to study the yields in $^{26}$Al derived from these
new rotating models (only the wind
contribution is considered here). 

\section{Effects of rotation on the $^{26}$Al yields from WR stars}

\begin{figure*}
\includegraphics[width=13cm,angle=0]{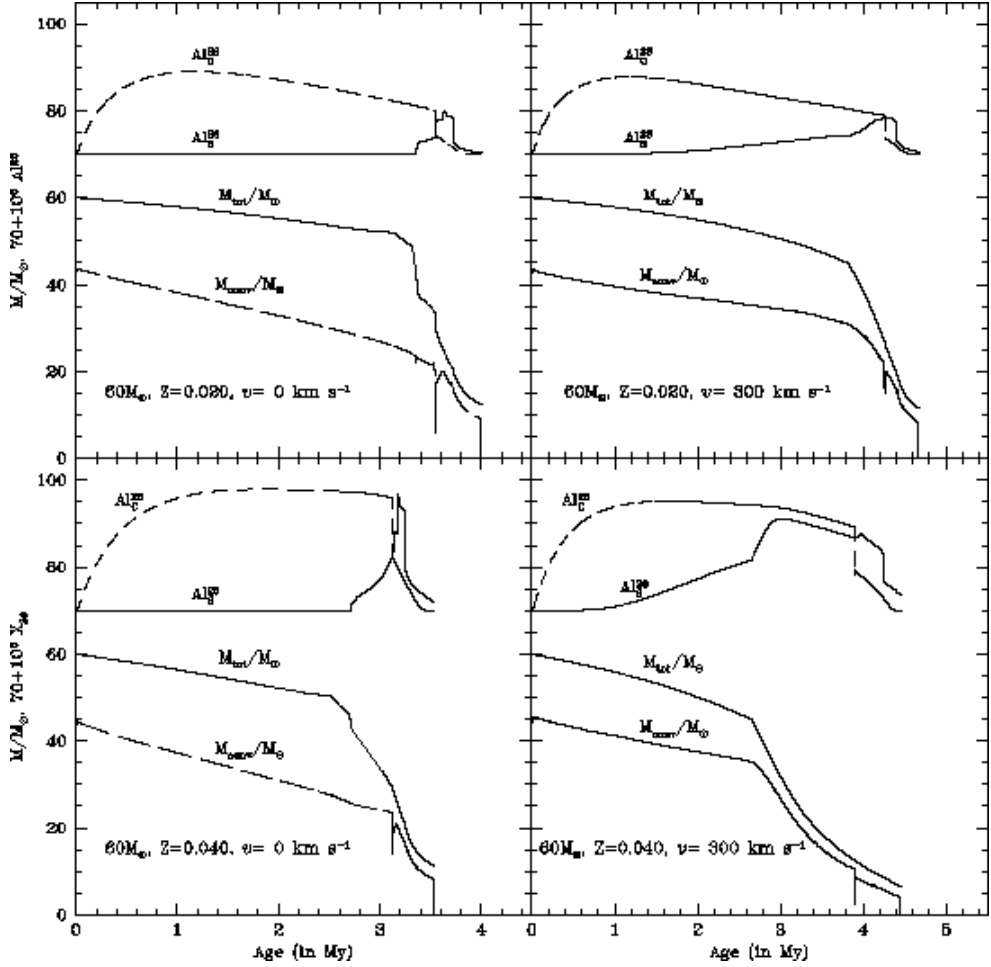}

\caption{Evolution as a function of time of the surface and central
abundance of $^{26}$Al in mass fraction
(labeled respectively by Al$^{26}_{\rm S}$ and Al$^{26}_{\rm C}$), of the total mass and of the mass
of the convective core in four 60 M$_\odot$ stellar models. The initial metallicities
and rotational velocities are indicated on each panel.}
\label{f1}
\end{figure*}

The physical ingredients of these models are exactly the same as 
in the models by Meynet \& Maeder \cite{ref1},
except for one thing: we did not take into account the effects of the wind anisotropies
induced by rotation. This is quite justified since it has
been shown in the above reference that for the initial velocities considered here
the effects are negligible. We consider here an initial equatorial
rotation velocity on the ZAMS of 300 km s$^{-1}$. This corresponds to a time
averaged equatorial velocity during the Main Sequence phase of about
200 km s$^{-1}$, a typical value for this kind of star. 
Figure \ref{f1} presents the evolution of the structure of four 60 M$_\odot$
stellar models with different metallicities and initial rotational velocities.
The evolutions of the central and surface abundances of $^{26}$Al are also
indicated. Comparing the non--rotating models with the rotating ones, one can note two
striking differences:
\begin{itemize}
\item  As a result of rotational mixing
the convective core in the rotating model
is continuously supplied with fuel and 
CNO elements which act as catalysts of the H--burning reactions.
Therefore, the
decrease in mass of the convective core is slowed down and the lifetime on the Main
Sequence is enhanced.
This feature is more marked in the
higher metallicity model, in which the greater CNO content
reinforces the process just described above. This effect
of rotation tends to favour $^{26}$Al production by WR stars.
Indeed more extended
convective cores reduce the extension of the region separating
the core, where $^{26}$Al is produced, from the surface, where it is ejected
into the interstellar medium by stellar winds.
\vskip 0.5mm
\item In the non--rotating models, the abundance in $^{26}$Al begins to increase
at the surface
when layers having been processed in the convective core appear at the surface
as a result of mass loss. In the rotating model, the enrichment in $^{26}$Al
of the surface occurs at a much earlier stage
when the core is still surrounded by an important H--rich envelope. 
This may happen thanks to rotational mixing which brings up to
the surface matter processed in the core. Thus
the wind becomes $^{26}$Al enriched at a much earlier stage than 
in the non--rotating models and the total quantity of ejected $^{26}$Al 
increases.
\end{itemize}

In addition to these two effects which are well apparent in Fig.~\ref{f1}, at least
two other effects contribute to enhance the $^{26}$Al yields in rotating models: 
1) some amounts of 
$^{25}$Mg diffuse from the radiative envelope in the central H--burning regions
where it is transformed by proton capture in $^{26}$Al, 2) the minimum
initial mass of single stars passing through a WR episode is decreased. For
all these reasons, 
the yields of the rotating models are enhanced with 
respect to those of non--rotating models computed with exactly the same
physical ingredients. The enhancement factors are slightly superior to 2 for the 60
and the 120 M$_\odot$ and amounts to 340 in the case
of the 25 M$_\odot$ stellar model. Note that the non--rotating
25 M$_\odot$ never becomes a WR star, while its rotating counterpart
goes through a WR phase (hence the huge enhancement factor).

\begin{figure*}
\includegraphics[width=8cm,angle=-90]{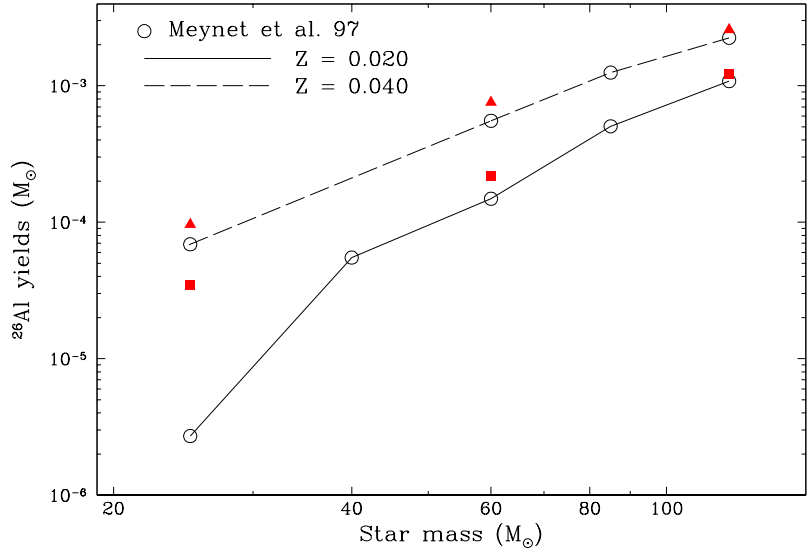}

\caption{Comparison of the $^{26}$Al yields obtained from the present rotating
models, filled symbols, with those of Meynet et al. \cite{ref7}, lines with open 
circles. Squares show the results for the rotating solar metallicity
models, and triangles for the rotating models with twice the solar 
metallicity.}
\label{f2}
\end{figure*}

On Fig.~\ref{f2} the yields from the present rotating models are compared with those
obtained from the enhanced mass loss rate models. 
One sees that the yields obtained in the new
rotating models, both at solar metallicity and at twice the solar metallicity,  
are greater than those of Meynet et al. \cite{ref7}, the increase becoming
more and more important when the initial mass decreases.
More quantitatively, at solar metallicity,  the yields for
the rotating 25, 60 and 120 M$_\odot$ stellar models are multiplied by a factor 12.7,
1.5 and 1.1 with respect to the yields of Meynet et al \cite{ref7} for the same masses.
In order to give an idea of the effects of these enhancement factors on the quantity
of $^{26}$Al ejected by a generation of stars, one can define 
an average yield, $\overline Y_{26}$, by $\overline Y_{26}={\int_{M_{\rm min}}^{M_{\rm max}} Y_{26} (M) \phi(M) {\rm d} M
\over \int_{M_{\rm min}}^{M_{\rm max}}  \phi(M) {\rm d} M}$, where $M_{\rm min}$, $M_{\rm max}$
are the lower and upper bound respectively of the mass range considered, $Y_{26}(M)$ the yield
in $^{26}$Al of the WR star of initial mass $M$ and $\phi(M)$ the initial mass function (IMF)
chosen here with a Salpeter slope (1.35).
The value of this average yield computed in the mass range between 25 and 70 M$_\odot$
is equal to $5.1\times 10^{-5}M_\odot$ when the stellar yields from the
enhanced mass loss rate models are used.
Computing the
same average yield from the rotating model, one obtains a value of $9.7\times 10^{-5}M_\odot$,
nearly twice as great as the one obtained from the enhanced mass loss rate models.

We mentionned above that observations of the Cygnus associations suggest that
the theoretical $^{26}$Al yields might be
underestimated by about a factor two. Indeed, Kn\"odlseder et al. \cite{ref5} converting
the observed 1.8 MeV flux measurements into an equivalent O7V star $^{26}$Al yield,
obtain a value $Y^{O7V}_{26} =(1.1\pm0.3)\times 10^{-4}M_\odot$
for the Cygnus region. This empirical mean yield per equivalent O7V star 
is about a factor two above the value
deduced from their population synthesis model ($Y^{O7V}_{26} =4.7\times 10^{-5}M_\odot$)
which, on the other hand, very well reproduces the ionizing luminosity of the
Cygnus region. 
How would the use of the present models change the situation ?
An appropriate quantitative assessement
of this point requires a careful study that will be made in a forthcoming paper.
Let us simply mention here that if rotation would only affect the $^{26}$Al yields, then
the above difficulty would be greatly alleviated by the use of the present yields.
However, rotation makes also the tracks slightly bluer and more luminous and
thus rotation will also increase the ionising luminosity of the stars. Thus the definite
answer to the above question remains largely open as long as no detailed
population syntheis models are performed.

What is the fraction of the 2-3 M$_\odot$ of $^{26}$Al believed to be present in the
Milky--Way which has been ejected by the WR stellar winds ?
On the basis of the enhanced mass loss rate
models the WR contribution was estimated to be  between 0.9 and 1.5 M$_\odot$
depending on the value of the slope of the IMF (respectively
1.7 and 1.35).
First estimates, using the present yields (Meynet et al. in preparation)
indicate that the contribution of the WR stars 
might amount to values between 1.2 (1.7) and 2 M$_\odot$ (1.35).
Thus, rotation  appears to reinforce the WR contribution to the observed 
$^{26}$Al in the Galaxy.

\section{Conclusion}

From the considerations above, we can retain the following conclusions:
\begin{itemize}
\item Rotation  increases the contribution of the WR stars to the $^{26}$Al
synthesis. A significant part of the increase results from the lowering
of the minimum initial mass of the stars going through a WR episode.
\item The increase of the $^{26}$Al yields due to rotation appears 
to be sufficient for significantly reducing
the difficulty encountered in reproducing the 1.8 MeV luminosity of the Cygnus region.
However it remains to see if these models can also reproduce the ionising
luminosity observed from this region.
\end{itemize}

\end{document}